\documentclass[twocolumn,eqsecnum,superscriptaddress,showpacs]{revtex4}
\usepackage[dvips]{graphicx}
\begin{document}
\hyphenation{anti-fermion}
\title{  HISTORICAL MISTAKE IN SPONTANEOUS SYMMETRY BREAKING  }
\vspace{1cm}
\author{T. Fujita}\email{fffujita@phys.cst.nihon-u.ac.jp}
\affiliation{Department of Physics, Faculty of Science and Technology, 
Nihon University, Tokyo, Japan}
\author{M.Hiramoto}\email{hiramoto@th.phys.titech.ac.jp}
\affiliation{Department of Physics, Faculty of Science, Tokyo Institute 
of Technology, Tokyo, Japan}
\author{H. Takahashi}\email{htaka@phys.ge.cst.nihon-u.ac.jp}
\affiliation{Department of Physics, Faculty of Science and Technology, 
Nihon University, Tokyo, Japan}

\date{\today}%

\begin{abstract}

The prediction of a massless boson after the spontaneous chiral  
symmetry breaking in the classic paper of Nambu and Jona-Lasinio (NJL) 
is shown to be wrong. 
Their mistake is due to a carelessly employed perturbative vacuum 
when evaluating the boson mass by summing up one loop Feynman diagrams. 
With the proper symmetry broken vacuum, one obtains a finite boson mass 
depending on the coupling constant $G$.

\end{abstract}
\maketitle


The chiral symmetry in some of fermion field theory models is 
spontaneously broken in the vacuum state, and Nambu and Jona-Lasinio (NJL) \cite{q7} 
first showed that, after the chiral symmetry breaking, there appeared 
a massless boson in the NJL model. This was just consistent with 
the Goldstone theorem \cite{q1,q2} which was, however, meant originally for 
boson field theory models. 

In this Letter, we show that  the prediction of a massless boson 
after the spontaneous chiral  symmetry breaking in the paper 
of Nambu and Jona-Lasinio  is wrong. 
Their mistake is due to a carelessly employed perturbative vacuum 
when evaluating the boson mass by summing up one loop Feynman diagrams 
where the Lagrangian density of the NJL model is written as
$$ {\cal L}= i \bar \psi  \gamma_{\mu} \partial^{\mu}  \psi  +G \left[ (\bar{\psi}\psi )^2 
+(\bar{\psi}i\gamma_5\psi )^2  \right]  . \eqno{(1)}  $$

Now, if one looks into the feature of the massless boson in the NJL paper, 
then one notices an unbelievable situation. That is, this massless boson 
should exist independent of the strength of the coupling constant $G$.  
Namely, the massless boson can be made out of fermions and 
antifermions even though the interactions between them are arbitrarily 
small or large. Since the coupling constant $G$ is the only parameter in 
the NJL model, this means that the massless boson exists from the beginning ! 
Of course this cannot be understood 
by any kind of normal dynamics. 

It is even more surprising that no physicist could ever raise 
any naive question on this abnormal dynamics. Why ? 
This is a Goldstone boson, and therefore can it be out of a clear sky ? 

But first we should clarify why Nambu and all other people 
obtained this abnormal massless boson. 
The answer is simple, but still it may take a few paragraphs to explain. 
The Lagrangian density has a chiral symmetry, and 
therefore one starts from the symmetric vacuum. This symmetric vacuum is 
taken to be the same as the perturbative vacuum and we denote it 
by $ |0\rangle $. The important discovery of Nambu and Jona-Lasinio was that 
there is a new vacuum whose energy is lower than the symmetric 
vacuum state. In addition, it breaks the chiral symmetry.  Therefore, 
we denote the new vacuum by $ |\Omega \rangle $. 
In general, quantum field theory is defined only when the vacuum is specified.  
In this case, one can develop calculation techniques  
like Feynman rules, and one obtains physical observables 
by calculating Feynman diagrams and so on. 
In the NJL model, it is clear that the physical vacuum state is $ |\Omega \rangle $ 
since the new vacuum energy is lower than the perturbative vacuum energy, 
and indeed that is what Nambu claimed.  

Now, the mistake Nambu and other people made in their calculations is simply 
that they calculated the boson mass by summing 
up one loop Feynman diagrams without paying attention for which of the vacuum states 
their formulation is based on. Indeed, they were based 
on the perturbative vacuum  $ |0\rangle $ simply because 
they just followed normal perturbation calculations. 

But here after the spontaneous symmetry breaking, that is just what 
should not be done. Since the perturbative vacuum  
$ |0\rangle $ is a false vacuum, 
the calculated boson mass was not a physical observable. Therefore, 
the boson mass they obtained could not depend on any thing. 
It is just there independent of the coupling constant $G$. 
As long as one evaluates the boson mass based on the false vacuum state, 
one inevitably ends up with a massless boson regardless their approximate 
schemes.  Therefore, the existence of this abnormal massless boson 
in the NJL model was well confirmed by any kind of calculations of 
many people \cite{q07}. 

Now, if one carries out the calculation of the boson mass by 
summing up one loop Feynman diagrams by the formulation based on the physical 
vacuum state $ |\Omega \rangle $, then one obtains a finite boson mass 
which indeed depends on the coupling constant $G$ or the approximate 
scheme one employs \cite{q9,q91}. For the vacuum state  $ |\Omega \rangle $ 
as determined by Nambu with the Bogoliubov transformation method, 
one obtains a finite boson mass for some regions of the coupling 
constant strength. But as one sees, the boson mass dependence 
on the coupling constant $G$ is a minor importance for the present discussion.  
Even more, there is a good reason to believe that there may well be 
no boson in the NJL as will be discussed later. 
Clearly, the important point is that one should start 
his field theory calculation based on the physical vacuum state  
even though it is more complicated and cumbersome than the evaluation 
with the perturbative vacuum state. 

There is one more reason why people overlooked the abnormality of 
the massless boson in the NJL calculation. That is due to the Goldstone theorem. 
The Goldstone theorem states that there should appear a massless boson 
after the spontaneous symmetry breaking. The Goldstone theorem is 
a very nice theorem which holds perfectly for boson field theory models. 
However, since the theorem was originally meant for the boson field theory, 
there is a very serious $ad \ hoc$ ansatz necessary for 
fermion field theory models. That is, the existence of the boson field 
$\phi$ itself which should be described by fermion field operators. 
This is nontrivial, and even in the two dimensional field theory model, 
the boson field $\phi$ cannot be normally described by fermion field operators. 
The only exception must be the Schwinger model where the boson field 
can be expressed by the fermion field operators \cite{m1}. 

Below, we briefly sketch the essence of the Goldstone theorem. 
The Goldstone theorem  starts from  
the existence of the vacuum expectation value of the following commutation relation, 
$$ \langle \Omega \mid [Q_5(t), \phi (0)] \mid \Omega \rangle  \neq 0  
\eqno{(2)} $$
where $Q_5(t)$ is a conserved charge associated with the symmetry and here the chiral charge. 
But the most important assumption in eq.(2) is the existence 
of the boson field $\phi (x)$ itself 
which must be described by the fermion fields, since otherwise, 
one cannot evaluate the commutation relation. 
Obviously, in the boson field theory models, 
this cannot be any problems since there are boson fields from the beginning, and 
the only thing that should be determined from eq.(2) is the energy dispersion 
of the boson field.  

If one can confirm the validity of eq.(2) together with the translational 
property,
$$ \phi (x) = e^{ipx} \phi (0) e^{-ipx} , \eqno{(3)} $$ 
 then  there should appear 
a massless boson after the spontaneous symmetry breaking. But this is just 
the tautology for the fermion field theory models. 
Namely, one wants to prove the existence of $\phi$ which 
eventually corresponds to a massless boson, but the existence of $\phi$ 
that can be described by fermion field operators has to be assumed. 
This is essentially different from the boson field theory where 
the dispersion relation of the momentum and energy 
is the only concern  for this field $\phi$. 

Here, we show a common mistake undertaken in the textbook to describe how 
the Goldstone theorem holds in fermion field theory model. One says 
that one may take the following $\phi_0$ as an example
$$  \phi_0(x) = \bar \psi(x) \gamma_5 \psi(x) . \eqno{ (4)}   $$
In this case, one can easily prove that this $\phi_0$ satisfies eq.(2). 
Therefore, it gives an impression that the Goldstone theorem is meaningful 
for the fermion field theory models as well. 

However, one easily notices that this $\phi_0$ has nothing to do with 
any bound state of fermions and antifermions. It only says that this $\phi_0$ 
has a fermion number zero operator. In this case, the statement one obtains 
from eq.(2) with the help of eq.(3) is, at most, that there should be a state 
in which fermion and antifermion have zero momentum and zero energy. 
This energy dispersion looks like 
a massless boson, but of course it has nothing to do with the massless boson. 

Obviously, the existence of the boson field $\phi$ can be confirmed only after 
the whole dynamics of this field theory model is completely solved. 
Further, even if one could solve the dynamics properly, it would not mean that 
the  $\phi$ can be expressed in terms of fermion field operators. 

Below, we show the Thirring model as an example of a fermion field theory model 
where there exists no boson field $\phi$, namely, 
no bound state of fermions and antifermions. In this case, 
it is obvious that eq.(2) is not satisfied.  

The Thirring model is exactly solved by the Bethe ansatz \cite{b5}. Further, it 
is almost a two dimensional version of the NJL model and therefore one may understand 
the situation in the NJL in a better way. 

Here, we summarize the result of the recent study by solving the Bethe ansatz equations 
analytically \cite{p5}. 
In \cite{p5}, the new vacuum state is discovered and the energy of the new vacuum 
is indeed lower than the well known symmetric vacuum state. The new vacuum state 
breaks the chiral symmetry.  The spontaneous symmetry breaking occurs in two dimensions 
in contradiction with Coleman's theorem \cite{q3}. The excitation spectrum  
has a finite gap and then starts the continuum states. This means that there is 
no boson field  $\phi$, and therefore there is no way to satisfy eq.(2), or 
eq.(2) itself does not make sense. Here, we should note that Coleman's theorem 
is not applicable for the field theory model which has no boson after 
the spontaneous symmetry breaking. 

On the other hand, if one solves the Thirring model with the Bogoliubov transformation 
in the same way as Nambu and Jona-Lasinio but with the proper symmetry broken vacuum, 
then one obtains a massive boson \cite{q9,q91}. 
This suggests that the Bogoliubov approximation tends to overestimate 
the attraction between fermions and antifermions to form a boson. 
Since the Thirring model is quite similar to the NJL model in the structure 
as far as the Bogolubov method is concerned, we expect that there should be 
no boson in the NJL model even though the Bogoliubov method predicts a massive boson. 
This is also due to the fact that a bound state is in general 
more difficult to make in four dimensions than in two dimensions. 

But, for the spontaneous symmetry breaking business, the existence of a massive 
boson is not a very important issue.



\vspace{2cm}

\end{document}